\documentclass[english,pra,twocolumn,groupedaddress,reprint]{revtex4}
\usepackage{graphicx}
\usepackage{amsfonts}
\usepackage[T1]{fontenc}
\usepackage[latin9]{inputenc}
\usepackage{color}
\usepackage{amsmath}
\usepackage{amssymb}
\usepackage{esint}
\usepackage{comment}
\usepackage{bbold}
\usepackage{tikz}
\usepackage{minitoc}
\usepackage{bm}
\usepackage{CJK}

\begin{document}

\title{Weak measurement amplification in optomechanics via a squeezed
coherent state pointer}
\author{Gang Li\footnote{ligang0311@sina.cn},$^{1}$ Li-Bo Chen,$^{2}$ Xiu-Min Lin,$^{3}$ and He-Shan
Song\footnote{hssong@dlut.edu.cn}$^{1}$}
\affiliation{$^{1}$School of Physics and Optoelectronic Technology, \\
Dalian University of Technology, Dalian 116024, China}
\affiliation{$^{2}$School of Science, Qingdao Technological University, Qingdao 266033,
China}
\affiliation{$^{3}$Fujian Provincial Key Laboratory of Quantum Manipulation and New
Energy Materials, \\
College of Physics and Energy, Fujian Normal University, Fuzhou 350007, China}
\date{\today }

\begin{abstract}
We present a scheme for achieving amplification of the displacement of the
mirror in optomechanical cavity using single-photon postselection where the
mirror is initially prepared in squeezed coherent state. The amplification
depends on the enhanced fluctuations of the squeezed coherent state, and it
is is caused by the noncommutativity of quantum mechanics relying on the
squeezed coherent state, which can not be explained by the standard weak
measurement \cite{Aharonov88,Simon11}.~~\newline
~~~~\newline
\end{abstract}

\maketitle

\section{Introduction}

Weak measurement theory \cite{Aharonov88}, first introduced by Aharonov,
Albert and Vaidman, describes a measurement situation where the measured
system is weakly coupled to the measuring device. Such weak measurement has
been used to solve basic problems in quantum mechanics \cite{Lundden09,Bamber11} and explain certain paradoxes \cite{Aharonov05}.

Weak measurement has been realized \cite{Ritchie91}, and can be applied to
amplifying tiny physical effects, such as spin Hall effect \cite{Hosten08}
and ultrasensitive beam direction \cite{Dixon09}, and measuring physical
quantity, such as the direct measurement of wave function \cite{Lundeen11}.
More experimental protocols have been proposed \cite{Romito08,Shpitalnik08,Brunner10,Zilberberg11,Wu12,Strubi13,Bula13,Meyer-Scott13,Dressel13}. Recently weak measurement protocol combined with cavity optomechanics \cite{Girvin09,Marquardt13} is given in \cite{Li14,Li-new}, and more applications of weak measurement are reviewed in \cite{Kofman12,Dressel14}.

In the most discussions about weak measurement, the initial state of the
pointer considered heretofore is Gaussian state (classical), such as \cite
{Aharonov88,Hosten08,Dixon09,Li14,Li-new}, but we are interested in whether
squeezed coherent state (quantum) pointer can further improve the
amplification effect of weak measurement compared to Gaussian pointer. In
this paper, we use the same optomechanical model in Ref. \cite
{Bouwmeester12,Li14} but the mirror is initially prepared in the squeezed
coherent state, and find that the amplification of the displacement of the
mirror depends on the enhanced fluctuations of squeezed coherent state,
which is lager than that with Gaussian pointer \cite{Li14,Li-new}. This
result is counter-intuitive since the enhanced fluctuations in a quadrature
of motion is considered to be of no use. To the best of our knowledge, this
is the first positive example for the use of the enhanced fluctuations of
the squeezed coherent state. The amplification effect is caused by the
noncommutativity of quantum mechanics relying on the squeezed coherent
state, which can not be explained by the standard weak measurement \cite
{Aharonov88,Simon11} (see Appendix A)
\begin{figure}[b]
\includegraphics[scale=0.38]{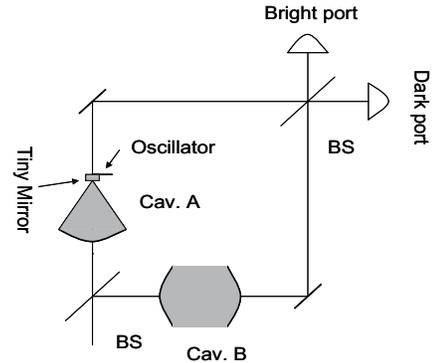}.
\caption{The photon enters the first beam splitter of March-Zehnder
interferometer with an optomechanical cavity A and a conventional cavity B.
The photon weakly affects the small mirror. After the second beam splitter,
dark port is detected, i.e., postselection acts on the case where the mirror
has been affected by a photon, and fails otherwise.}
\end{figure}

The structure of our paper is as follows. In Sec. II, we state the main
result of this work, including the amplification effect with squeezed vacuum
state and squeezed coherent state pointers, and In Sec. III, we give the
conclusion about the work.

\section{Amplification in optomechanics}

\subsection{The optomechanical model}

In Fig. 1 the optomechanical cavity A is embedded in one arm of the
March-Zehnder interferometer and a stationary Fabry-Peot cavity B is placed
in another arm. The two beam splitters are both symmetric. When the photon
is detected at the dark port, the dynamical evolution of the photon and the
movable mirror is finished, i.e., the orthogonal postselection is performed.
The Hamiltonian of optomechanical system is expressed as:
\begin{equation}
H=\hbar \omega _{0}(a^{\dagger }a+b^{\dagger }b)+\hbar \omega _{m}c^{\dagger
}c-\hbar ga^{\dagger }a(c^{\dagger }+c),  \label{a}
\end{equation}
where $\hbar $ is Plank's constant. $\omega _{0}$ and $a$ (or $b$) are
frequency and annihilation operator of the optical cavity A (or B),
respectively. $c$ is annihilation operator of mechanical system with angular
frequency $\omega _{m}$ and the optomechanical coupling strength $g=\frac{
\omega _{0}}{L}\sigma $, where $L$ is the length of the cavity A (or B). $
\sigma =(\hbar /2m\omega _{m})^{1/2}$ is the zero-point fluctuation and $m$
is the mass of mechanical system. Here it is a weak measurement model where
the mirror is used as the pointer to measure the number of photon in cavity
A, with $a^{\dag }a$ of the Eq. (\ref{a}) corresponding to $\hat{\sigma}_{z}$
in Eq. (\ref{sss}) in the standard scenario of weak measurement (see
Appendix A) and $c+c^{\dag}$ corresponds to $\hat{p}$.

If the initial state of the mirror can be prepared at the squeezing coherent
state $|\alpha ,\varepsilon \rangle =S(\varepsilon )|\alpha \rangle $, where
$S(\varepsilon )=\exp (-\frac{1}{2}\varepsilon c^{\dagger 2}+\frac{1}{2}
\varepsilon ^{\ast }c^{2})$ with squeezing parameters $\varepsilon
=re^{i\theta }$ and one photon interacts with the mirror, according to the
results of the Hamiltonian in \cite{Bose97,Mancini97}, the state of the
mirror will become
\begin{equation}
|\psi (\xi ,\eta ,\varphi )\rangle =e^{i\phi (t)}D(\xi (t))S(\eta
(t))|\varphi (t)\rangle ,  \label{b}
\end{equation}
where $e^{i\phi (t)}$ is the Kerr phase of one photon with $\phi
(t)=k^{2}(\omega _{m}t-\sin \omega _{m}t)$, $\varphi (t)=\alpha e^{-i\omega
_{m}t},\eta =re^{i(\theta -2\omega _{m}t)}$ and $D(\xi (t))=\exp [\xi
(t)c^{\dag }-\xi ^{\ast }(t)c]$ is a displacement operator with $\xi
(t)=k(1-e^{-i\omega _{m}t})$. The position displacement of the mirror caused
by one photon is $\langle \psi (\xi ,\eta ,\varphi )|\hat{q}|\psi (\xi ,\eta
,\varphi )\rangle -\langle \varphi (t)|S^{\dagger }(\eta (t))\hat{q}S(\eta
(t))|\varphi (t)\rangle $ with $\hat{q}$ being the position operator. It can
be shown that the displacement can not be bigger than $4k\sigma $ for any
time $t$. Since $k=g/\omega _{m}$ can not be bigger than $0.25$ in weak
coupling condition \cite{Marshall03}, then the position displacement of the
mirror caused by one photon can not be bigger than the zero-point
fluctuation $\sigma $. In the literature \cite{Marshall03} we know that if
the displacement of the mirror can be detected experimentally it should be
not smaller than $\sigma $. Therefore, the displacement of the mirror caused
by one photon can not be detected. In the following we will show how the
weak measurement can amplify the mirror's displacement.

\subsection{Amplification about position variable $\hat{q}$ with a squeezed
vacuum state pointer}

Suppose that one photon is input into the interferometer, the state of the
photon after the first beam splitter becomes
\begin{equation}
|\psi _{i}\rangle =\frac{1}{\sqrt{2}}(|1\rangle _{A}|0\rangle _{B}+|0\rangle
_{A}|1\rangle _{B}),  \label{c}
\end{equation}
and after interacting weakly with the mirror prepared at the squeezing
vacuum state $S(\varepsilon )|0\rangle $, the state of the total system is
\begin{eqnarray}
|\psi (t)\rangle &=&\frac{1}{\sqrt{2}}(|1\rangle _{A}|0\rangle _{B}|\psi
(\xi ,\eta ,0)\rangle _{m}+|0\rangle _{A}|1\rangle _{B}  \notag \\
&&S(\eta )|0\rangle _{m}).  \label{d}
\end{eqnarray}
After the second beam splitter and detecting at dark port, that is,
postselecting for an single-photon state
\begin{equation}
|\psi _{f}\rangle =\frac{1}{\sqrt{2}}(|1\rangle _{A}|0\rangle _{B}-|0\rangle
_{A}|1\rangle _{B}),  \label{e}
\end{equation}
which satisfies this case of the orthogonal postselection, i.e., $
\langle\psi_{f}|\psi_{i}\rangle =0$. Then the final state of the mirror
becomes
\begin{equation}
|\Psi _{os}(t)\rangle =\frac{1}{2}(|\psi (\xi ,\eta ,0)\rangle _{m}-S(\eta
)|0\rangle _{m}).  \label{f}
\end{equation}

The average displacement of pointer variable $\hat{q}$ of the mirror is
\begin{eqnarray}
\langle \hat{q}\rangle &=&\frac{Tr(|\Psi _{os}(t)\rangle \langle \Psi
_{os}(t)|\hat{q})}{Tr(|\Psi _{os}(t)\rangle \langle \Psi _{os}(t)|)}
-Tr(S(\eta )|0\rangle _{m}\langle 0|_{m}  \notag \\
&&S^{\dagger }(\eta )\hat{q}).  \label{g}
\end{eqnarray}
Here we use the position operator $\hat{q}=(c+c^{\dagger})\sigma $.
Substituting Eq. (\ref{f}) into Eq. (\ref{g}), as a result, we have
\begin{eqnarray}
\langle q(t)\rangle &=&\sigma \lbrack \xi (t)+\xi ^{\ast }(t)-e^{-\frac{
|\upsilon (t)|^{2}}{2}}(e^{i\phi (t)}\upsilon (t)\mu ^{\ast }(t)  \notag \\
&+&e^{-i\phi (t)}\upsilon ^{\ast}(t)\mu (t))]/[2-e^{-\frac{|\upsilon
(t)|^{2} }{2}}(e^{i\phi (t)}  \notag \\
&+&e^{-i\phi (t)})],  \label{h}
\end{eqnarray}
where $\mu (t)=\cosh r-e^{i(\theta -2\omega _{m}t)}\sinh r$ and $\upsilon
(t)=\xi (t)\cosh r+\xi ^{\ast }(t)e^{i(\theta -2\omega _{m}t)}\sinh r$.

The average displacement $\langle q(t)\rangle /\sigma $ of the mirror is
shown in Fig. 2 as a function of $\omega _{m}t$ with $k=0.005$, $r=2$ and $
\theta =\pi $. We can see that there are two amplification zones. One is
that there exist amplification effects around time $\omega _{m}t=(2n+1)\pi $
$(n=0,1,2,\cdots )$. The other amplifications occur around the vibration
periods of the mirror $\omega _{m}t=2n\pi $ $(n=1,2,\cdots )$. In
particular, the maximal positive and negative amplifications around time $
\omega _{m}t=2n\pi $ $(n=1,2,\cdots )$ are more prominent and they can also
reach the maximal value \cite{Marshall03} $\langle q\rangle =\pm e^{2}\sigma
$, i.e., the level of the squeezing vacuum-state fluctuation $\pm \sqrt{
e^{2r}\sin ^{2}\frac{\theta }{2}+e^{-2r}\cos ^{2}\frac{\theta }{2}}\sigma $
when $r=2$ and $\theta =\pi $ ($\sigma $ the level of the vacuum-state
fluctuation). Note that the maximal displacement of the mirror caused by one
photon in cavity A (see Fig. 1) is $4k\sigma $, therefore the amplification
factor is $Q=\pm e^{2}/4k$ which is $\pm 369.5$ when $k=0.005$. Compared
with Ref. \cite{Li14}, not only the maximal amplifications around time $
\omega _{m}t=2n\pi $ $(n=1,2,\cdots )$ are generated, but also the maximal
amplification can be much larger since the squeezing vacuum state of the
mirror is prepared.

\begin{figure}[t]
\includegraphics[scale=0.56]{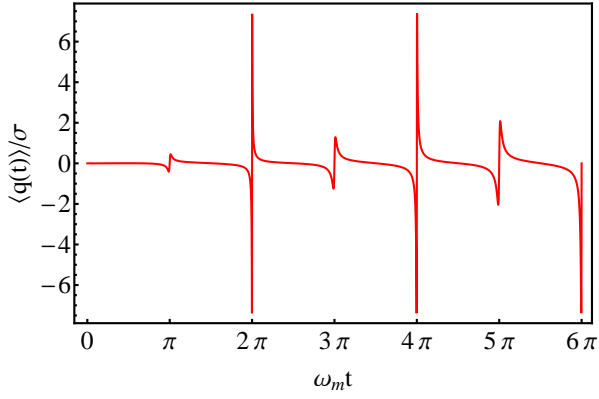}
\caption{The average displacement $\langle q(t)\rangle/\protect\sigma$ as a
function of $\protect\omega_{m}t$ with $k=0.005$, $r=2$ and $\protect\theta=
\protect\pi$.}
\end{figure}

In order to understand the amplification effects appearing around time $
T=2n\pi $ $(n=1,2,\cdots )$, we can perform a small quantity expansion about
time $T$ till the second order when $\theta =\pi $. Suppose that $|\omega
_{m}t-T|\ll 1$, $k\ll 1$ and $k^{2}T\ll 1$, then
\begin{eqnarray}
|\Psi _{os}(t)\rangle &\approx &\frac{1}{2}[ik^{2}TS(\eta _{1})|0\rangle
_{m}+ik(\omega _{m}t-T)  \notag \\
&&e^{r}S(\eta _{1})|1\rangle _{m}],  \label{aa}
\end{eqnarray}
where $\eta _{1}=re^{i(\pi -2(\omega _{m}t-T))}$. It should be noted that we
use the approximation $e^{i\phi (t)}\approx 1+ik^{2}T$. It can be easily
seen that the squeezed vacuum state $S(\eta _{1})|0\rangle $ of Eq. (\ref{aa}
) is generated due to the kerr phase. Substituting Eq. (\ref{aa}) into Eq. (
\ref{g}), the average value of displacement operator $\hat{q}$ is given by
\begin{eqnarray}
\langle q(t)\rangle &=&\sigma 2k^{3}Te^{2r}(\omega
_{m}t-T)/[k^{4}T^{2}+k^{2}(\omega _{m}t  \notag \\
&&-T)^{2}e^{2r}],  \label{bb}
\end{eqnarray}
which gets its maximal value $e^{r}\sigma $ when $k^{2}T=k(\omega
_{m}t-T)e^{r}$, and gets its minimal value $-e^{r}\sigma $ when $
k^{2}T=-k(\omega _{m}t-T)e^{r}$. Therefore, the mirror state achieving the
positive amplification is $S(\eta _{1})\frac{1}{\sqrt{2}}(|0\rangle
_{m}+|1\rangle _{m})$ and the state achieving the negative amplification is $
S(\eta _{1})\frac{1}{\sqrt{2}}(|0\rangle _{m}-|1\rangle _{m})$. It is
obvious that the key to understand the amplification is the superposition of
the squeezing vacuum state and the squeezing one-phonon state of the mirror,
which is due to the Kerr phase. It is well-known that the squeezed state
have many application in the diverse fields of quantum physics since it have
less fluctuations in one quadrature component at the expense of increased
fluctuations in another quadrature component, which is a key characteristic
of the squeezed state on the application side. On the contrary, the
application of its quadrature component with increased fluctuations is
seldom mentioned. Here we show that the amplification achieved here is
associated with the quadrature component with increased fluctuations in the
squeezed state, i.e., the level of the squeezing vacuum-state fluctuation $
\pm e^{r}\sigma $. We can see that the amplification will become bigger if
the level of the increased fluctuations of squeezed coherent state in
another quadrature direction become larger and it becomes to be measured
more easily. It will be shown that the maximal amplification due to the Kerr
phase obtained here is bigger than the level of vacuum-state fluctuation $
\sigma $ which can't be achieved if the mirror is initially prepared in the
ground state \cite{Li14} or the coherent state \cite{Li-new}. And in this
case of the orthogonal postselection, the amplifications around time $\omega
_{m}t=(2n+1)\pi $ $(n=0,1,2,\cdots )$ or $\omega _{m}t=2n\pi $ $(n=1,2,\cdots )$ are not explained by the weak measurement \cite{Aharonov88,Simon11} (see Appendix A) which is applicable to the case that
is an near-orthogonal postselection.

\subsection{Amplification about position variable $\hat{q}$ with a squeezed
coherent state pointer}

Suppose that the mirror is initially prepared at the squeezed coherent state
$S(\varepsilon )|\alpha \rangle $ with $\alpha =|\alpha |e^{i\beta }$, where
$|\alpha |$ and $\beta $ are real numbers called the amplitude and phase of
the state, respectively. If steps corresponding to those of Eqs. (\ref{c})--(
\ref{e}) are carefully carried out in this case, then the final state of
mirror becomes
\begin{equation}
|\Psi _{os}(t)\rangle =\frac{1}{2}(|\psi (\xi ,\eta ,\varphi )\rangle
_{m}-S(\eta (t))|\varphi (t)\rangle _{m}).  \label{cc}
\end{equation}
For the sake of making the analysis simple, we can displace the state of Eq.
(\ref{cc}) to the origin point in phase space, defining $|\chi
_{os}(t)\rangle =D^{\dag }[e^{-i\omega _{m}t}(\alpha \cosh r-\alpha ^{\ast
}e^{i\theta }\sinh r)]|\Psi _{os}(t)\rangle $ and there is

\begin{equation}
|\chi _{os}(t)\rangle =\frac{1}{2}(e^{i\phi (t)+i\tau }D(\xi (t))S(\eta
)|0\rangle _{m}-S(\eta )|0\rangle _{m}),  \label{dd}
\end{equation}
where the phase $e^{i\tau }$ with $\tau (t)=-i[\alpha ^{\ast }\upsilon
e^{i\omega _{m}t}-\alpha \upsilon ^{\ast }e^{-i\omega _{m}t}]$ is a relative
phase between the states $D(\xi (t))S(\eta )|0\rangle _{m}$ and $S(\eta
)|0\rangle _{m}$ and it is obtained when we use the property of the
displacement operators $D(\alpha )D(\beta )=\exp [\alpha \beta ^{\ast
}-\alpha ^{\ast }\beta ]D(\beta )D(\alpha )$, i.e., the noncommutativity of
quantum mechanics relying on the squeezed coherent state, which is similar
to the derivation of the relative phase caused by the coherent state in \cite
{Li-new}.

Substituting Eq. (\ref{dd}) into Eq. (\ref{g}), as a result, we have
\begin{eqnarray}
\langle q(t)\rangle &=&\sigma \lbrack \xi (t)+\xi ^{\ast }(t)-e^{-\frac{
|\upsilon (t)|^{2}}{2}}(e^{i\phi (t)+i\tau (t)}\upsilon (t)\mu ^{\ast }(t)
\notag \\
&+&e^{-i(\phi (t)+\tau (t))}\upsilon ^{\ast }(t)\mu (t))]/[2-e^{-\frac{
|\upsilon (t)|^{2}}{2}}(e^{i\phi (t)}  \notag \\
&&e^{i\tau (t)}+e^{-i(\phi (t)+\tau (t))})].  \label{ee}
\end{eqnarray}

\begin{figure}[t]
\includegraphics[scale=0.45]{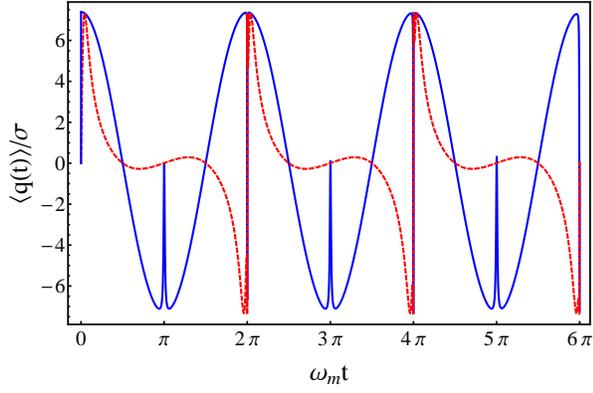}
\caption{The average displacement $\langle q(t)\rangle/\protect\sigma$ as a
function of $\protect\omega_{m}t$ with $k=0.005$, $r=2$ and $\protect\theta=
\protect\pi$ for the different amplitudes and the phases, $|\protect\alpha
|=1/2$, $\protect\beta=2\protect\pi$ (solid line) and $|\protect\alpha|=400$
, $\protect\beta=\protect\pi/2$ (dashed line).}
\end{figure}

In Fig. 3, the average displacement $\langle q(t)\rangle /\sigma $ of the
mirror versus time $\omega _{m}t$ with $k=0.005$ $r=2$ and $\theta =\pi $
for different amplitudes and the phases, $|\alpha |=1/2$, $\beta $ $=0$
(solid line) and $|\alpha |=400$, $\beta $ $=\pi /2$ (dashed line). It can
be seen clearly that the maximal amplification can not only occur around $
\omega _{m}t=(2n+1)\pi $ $(n=0,1,\cdots )$ and the vibration periods of the
mirror $\omega _{m}t=2n\pi $ $(n=1,2,\cdots )$ but also occur near the
initial time where their maximal values are $e^{2}\sigma $, i.e., the level
of the increased fluctuations of the squeezed coherent state.

Similar to Eq. (\ref{aa}), we can also perform a small quantity expansion
about time $T=0$ till the second order when $\theta =\pi $. Suppose that $
|\omega _{m}t-T|\ll 1$, i.e., $\omega _{m}t\ll 1$, and $k\ll 1$, then
\begin{equation}
|\chi _{os}(t)\rangle \approx \frac{1}{2}[i2k|\alpha |\zeta S(\eta
_{2})|0\rangle _{m}+ik\omega _{m}te^{r}S(\eta _{2})|1\rangle _{m}],
\label{ff}
\end{equation}
where $\eta _{2}=re^{i(\pi -2\omega _{m}t)}$ and $\zeta =e^{r}\omega
_{m}t\cos \beta +e^{-r}(\omega _{m}t)^{2}\frac{\sin \beta }{2}$. Noted that
we use the approximation $e^{i\tau }\approx 1+i2k|\alpha |\zeta $. It can be
easily seen that the squeezed vacuum state $S(\eta _{2})|0\rangle $ of Eq. (
\ref{ff}) is generated due to the relative phase $e^{i\tau }$. Substituting
Eq. (\ref{ff}) into Eq. (\ref{g}), the average value of displacement
operator $\hat{q}$ is given by
\begin{equation}
\langle q(t)\rangle =\sigma 4k^{2}|\alpha |\zeta e^{2r}\omega
_{m}t/[4k^{2}|\alpha |^{2}\zeta ^{2}+k^{2}(\omega _{m}t)^{2}e^{2r}],
\label{gg}
\end{equation}
which gets its maximal value $e^{r}\sigma $ or minimal value $-e^{r}\sigma $
when $2k|\alpha |\zeta =\pm ke^{r}\omega _{m}t$, respectively. Results
clearly indicate that the maximal amplifications appearing near the initial
time arise from the equal superposition of squeezing vacuum state and the
squeezing one-phonon state of the mirror, i.e., $S(\eta _{2})\frac{1}{\sqrt{2
}}(|0\rangle _{m}\pm |1\rangle _{m})$, which is due to the relative phase $
e^{i\tau }$ caused by the noncommutativity of quantum mechanics.

As a result, the amplifications due to the phase $e^{i\tau }$ occurring near
the initial time is different from the amplifications due to the Kerr phase
since the reasons for the two amplifications are essentially different. The
relative phase $e^{i\tau }$ after an orthogonal postselection is caused by
the noncommutativity of quantum mechanics. In the standard weak measurement
\cite{Aharonov88,Simon11} the relative phase results from near-orthogonal
postselection (see Appendix A). Therefore, the amplification using the
squeezed coherent state can not be explained by the weak measurement \cite
{Aharonov88,Simon11}. Remarkably, the amplifications using the squeezed
coherent state not only depend on the quadrature component with increased
fluctuations but also can appear near the initial time, i.e., one photon is
successfully detected at dark port within a very short time. The maximal
amplification using the squeezed coherent state can reach the level of the
increased fluctuations $e^{r}\sigma $ and it can be detected more easily
than the maximal amplification value (the level of the vacuum-state
fluctuation $\sigma $) with the ground state pointer \cite{Li14} or the
coherent state pointer \cite{Li-new} since $e^{r}\sigma $ exceed the strong
coupling limiting $\sigma $ \cite{Marshall03}. Therefore, the amplification
result provided here is a surprising one in weak coupling regime.

Taking into account of dissipation, the master equation of the mechanical
system \cite{Bose97} is given by
\begin{eqnarray}
\frac{d\rho (t)}{dt} &=&-\frac{i}{\hbar }[H,\rho (t)]  \notag \\
&+&\frac{\gamma _{m}}{2}[{2c\rho (t)c^{\dag }-c^{\dag }c\rho (t)-\rho
(t)c^{\dag }c]},  \label{ii}
\end{eqnarray}
where $\gamma _{m}$ is the damping constant.

Similar to the result of the dissipation in Ref. \cite{Li14}, because the
actual $\gamma =\gamma _{m}/\omega _{m}$ can be very small ($\gamma =5\times
10^{-7}$ in Proposed device no. 2 \cite{Bouwmeester12}) and is almost close
to $\gamma =0$, all the amplification values in the presence of the damping
are almost unscathed.

\begin{figure}[t]
{\includegraphics[scale=0.47]{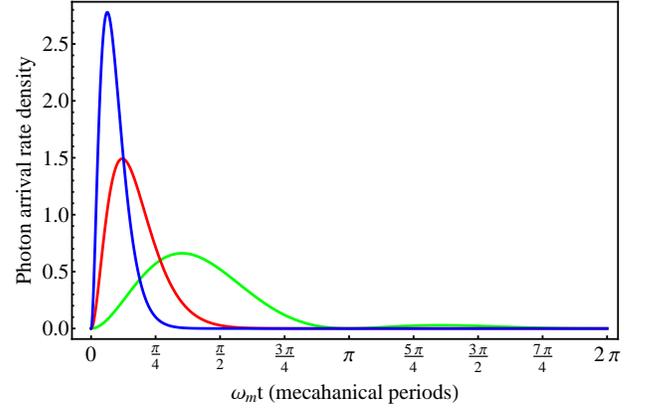}}
\caption{For $|\protect\alpha|=1/2$, $\protect\beta=2\protect\pi$ and $r=2$,
$\protect\theta=\protect\pi$, Photon arrival probability density vs arrival
time for a successful postselection with $\protect\kappa=\protect\omega_{m}$
(green line), $\protect\kappa=5\protect\omega_{m}$ (red line) and $\protect
\kappa=10\protect\omega_{m}$ (blue line)}
\end{figure}

\subsection{Discussion}

Considering the feasibility of the proposed scheme, we discuss the
experimental requirements. First, the mechanical oscillator (mirror) of our
device is initially prepared in a squeezed coherent state. Coherent state of
the mechanical oscillator has been prepared with itinerant microwave fields
\cite{Lehnert13}. Achieving squeezed states in mechanical oscillators has
not been realized experimentally but proposed by several schemes, such as
resolved sideband cooling using squeezed \cite{Jahne09} or modulated input
light \cite{Mari09}.

It is difficult to observe the amplification effect using the squeezed
vacuum state pointer because of the very small post-selected probability of
success and the very limited time zones for the appearance of the maximal
displaced state. Nevertheless, the amplification using the squeezed coherent
state pointer is not the case. The probability density of a photon released
from an optomechanical cavity after time $\omega _{m}t$ is $\kappa \exp
(-\kappa t)$, where $\kappa $ is the decay rate of the cavity. The
probability of a successful postselection released after $\omega _{m}t$ is $
\frac{1}{4}[2-e^{-\frac{|\upsilon (t)|^{2}}{2}}(e^{i\phi (t)+i\tau
(t)}+e^{-i(\phi (t)+\tau (t))})]$. For $k\ll 1$, this is approximately $
\frac{1}{4}[|\upsilon (t)|^{2}+\tau (t)^{2}]$. Multiplying these results,
the photon arrival rate density in an optomechanical cavity will be given by
\begin{equation}
\frac{\kappa }{4P}\exp (-\kappa t)(|\upsilon (t)|^{2}+\tau (t)^{2}),
\label{jjk}
\end{equation}
where $P$ is overall single photon probability of the state in Eq. (\ref{dd}
):

\begin{eqnarray}
P &=&\frac{1}{4}\int_{0}^{\infty }\kappa \exp (-\kappa t)(|\upsilon
(t)|^{2}+\tau (t)^{2})dt  \notag \\
&=&\frac{k^{2}\kappa (3\omega _{m}^{4}+2e^{8}\omega _{m}^{4}+e^{8}\omega
_{m}^{2}\kappa ^{2})}{2e^{4}(\kappa ^{5}+5\kappa ^{3}\omega _{m}^{2}+4\kappa
\omega _{m}^{4})}  \label{jj}
\end{eqnarray}%
when $|\alpha |=\frac{1}{2},\beta =2\pi ,r=2$ and $\theta =\pi $. The photon
arrival rate density is shown in Fig. 4. It can be seen clearly that in the
bad-cavity limit $\kappa >\omega _{m}$, i.e., non-sideband resolved regime,
and as the decay rate of the cavity $\kappa $ increases, such as $\kappa
=10\omega _{m}$, the photon arrival rate density increasingly distributes
mainly at time $t$ near $0$ where it is very narrow. Because of the photon
arrival rate concentrating near the zero time (blue line) in Fig. 4 and the
maximal amplification occurring at time $t$ near $0$ (solid line) in Fig. 3,
for a repeated experimental setup with identical conditions, the "average"
position displacement of the pointer is given by

\begin{eqnarray}
\overline{\langle q(t)\rangle } &=&\frac{\kappa }{4P}\int_{0}^{\infty }\exp
(-\kappa t)(|\upsilon (t)|^{2}+\tau (t)^{2})\langle q(t)\rangle dt  \notag \\
&=&6.98\sigma ,  \label{jjj}
\end{eqnarray}
where $\langle q(t)\rangle $ is the same as $\langle q(t)\rangle $ in Eq. (
\ref{ee}). Noted that the result can be completely detected in experiment
since it is beyond the strong coupling limit \cite{Marshall03}, i.e., the
level of vacuum-state fluctuation $\sigma $.

Second, we discuss experimental requirements for the optomechanical device.
The successful postselection probability of our needed device is common,
though precise value of which depends on the dark count rate of the detector
and the stability of the setup. As shown in Eq. (\ref{jj}), the probability
of successful postselection in an optomechanical device with $\kappa
=10\omega _{m}$ is approximately $0.525k^{2}$. The window in which the
detectors will need to be open for photons is approximately $1/\kappa $,
leading to a requirement that the dark count rate be lower than $%
0.525k^{2}\kappa $. Because the best silicon avalanche photodiode have dark
count rate of $\sim 2$ Hz, so we get $k$ $\geq $ $0.0036$ for a $4.5$ kHz
device \cite{Bouwmeester12}, i.e., Proposed device no. 2, but $\kappa
=10\omega _{m}$. In other words, the optical finesse $F$ in Proposed device
no. 2 is reduced to $3.33\times 10^{4}.$ Such an optomechanical cavity is
easily to be prepared. Therefore, the implementation of the scheme provided
here is feasible in experiment.

\section{Conclusion}

In summary, when combined with weak measurement, we use the squeezed
coherent state to enhance the displacement of the mirror in optomechanical
system. The amplification of the displacement of the mirror depends on the
enhanced fluctuations of squeezed coherent state, which is larger than that
with Gaussian state \cite{Li14,Li-new}. Such result is due to the
noncommutativity of quantum mechanics relying on the squeezed coherent
state, which can't be explained by the standard weak measurement \cite{Aharonov88,Simon11}. Moreover, the amplification occurring at time near $t= 0$, which is important for bad optomechanical cavity with non-sideband
resolved regime, makes our proposed scheme feasible in experiment. These
results not only extend the application of weak measurement to
optomechanical system, but also deepen our understanding of the weak
measurement.

\section*{Acknowledgments}

This work was supported by by the National Natural Science Foundation of
China under grants No. 11175033. L. B. C. acknowledges support from the
National Natural Science Foundation of China (Grant No.11304174) and Natural
Science Foundation of Shandong Province (Grant No. ZR2013AQ010). X. M. L.
acknowledges support from the National Natural Science Foundation of China
(Grant Nos. 61275215) and the National Fundamental Research Program of China
(Grant No. 2011CBA00203).

\appendix

\section{ Weak measurement with a ground state pointer}

In Ref. \cite{Simon11}, they consider the standard weak measurement model
but the initial state of the pointer is assumed to be the ground state $
|0\rangle _{m}$. Suppose the state $|+\rangle =\frac{1}{\sqrt{2}}(|0\rangle
_{s}+|1\rangle _{s})$ is the initial state of the system to be measured,
where $|0\rangle _{s}$ and $|1\rangle _{s}$ is eigenstates of $\hat{\sigma}
_{z}$. The Hamiltonian between the pointer and the system is given in
general as
\begin{equation}
\hat{H}=\hbar \chi (t)\hat{\sigma}_{z}\otimes \hat{p},  \label{sss}
\end{equation}
where $\sigma _{z}$ is an observable of the system to be measured, $\hat{p}$
is the momentum operator of the pointer and $\chi (t)$ is a narrow pulse
function with integration $\chi $. Suppose $\hat{q}$ is position operator of
the pointer that is conjugates to $\hat{p}$, therefore there is $
[q,p]=i\hbar $. As in Ref. \cite{Simon11}, if defining an annihilation
operator $\hat{c}=\frac{1}{2\sigma }\hat{q}+i\frac{\sigma }{\hbar }\hat{p}$,
where $\sigma $ is the zero-point fluctuation of the pointer ground state,
the Hamiltonian of Eq. (\ref{sss}) can be rewritten as
\begin{equation}
\hat{H}=-i\frac{\hbar ^{2}\chi (t)}{2\sigma }\hat{\sigma}_{z}(\hat{c}-\hat{c}
^{\dagger }).  \label{ttt}
\end{equation}
Then the time evolution of the total system is given by
\begin{eqnarray}
e^{-\frac{i}{\hbar }\int \hat{H}dt}|+\rangle |0\rangle _{m} &=&\exp [-\eta
\hat{\sigma}_{z}(\hat{c}-\hat{c}^{\dagger })]|+\rangle |0\rangle _{m}  \notag
\\
&=&\frac{1}{\sqrt{2}}(|0\rangle _{s}D(\eta )|0\rangle _{m}  \notag \\
&+&|1\rangle _{s}D(-\eta )|0\rangle _{m}),  \label{ss}
\end{eqnarray}%
where $D(\eta )=\exp [\eta \hat{c}^{\dagger }-\eta ^{\ast }\hat{c}]$ with $
\eta =\frac{\hbar \chi }{2\sigma }$ is a displacement operator and $\eta \ll
1$. In the weak measurement regime \cite{Aharonov88} the post-selected state
of the system is closely orthogonal to the initial state of the system which
is usually chosen as $\varepsilon |+\rangle +|-\rangle $, where $
|\varepsilon |\ll 1$. After postselection the final state of the pointer
became
\begin{eqnarray}
|\psi \rangle _{m} &=&\frac{1}{\sqrt{2}}[(1+\varepsilon )D(\eta )|0\rangle
_{m}-(1  \notag \\
&&-\varepsilon )D(-\eta )|0\rangle _{m}].  \label{tt}
\end{eqnarray}
When $|\varepsilon |\ll 1$ and $\eta \ll 1$, there is
\begin{eqnarray}
|\psi \rangle _{m} &\approx &\frac{1}{2}[(1+\varepsilon )(1-\eta \hat{\sigma}
_{z}(\hat{c}-\hat{c}^{\dagger }))|0\rangle _{m}  \notag \\
&-&(1-\varepsilon )(1-\eta \hat{\sigma}_{z}(\hat{c}-\hat{c}^{\dagger
}))|0\rangle _{m}]  \notag \\
&\approx &\varepsilon |0\rangle +\eta |1\rangle .  \label{xx}
\end{eqnarray}
Noted that the tiny relative phase $\varepsilon $ arise from a
near-orthogonal postselection on the system. Using the expression of the
pointer's displacement
\begin{equation}
\langle \hat{q}\rangle =\frac{\langle \psi |_{m}\hat{q}|\psi \rangle _{m}}{
\langle \psi |\psi \rangle _{m}}-\langle 0|_{m}\hat{q}|0\rangle _{m},
\label{yy}
\end{equation}
and
\begin{equation}
\langle \hat{p}\rangle =\frac{\langle \psi |_{m}\hat{p}|\psi \rangle _{m}}{
\langle \psi |\psi \rangle _{m}}-\langle 0|_{m}\hat{p}|0\rangle _{m}.
\label{zz}
\end{equation}
Hence in this case of the near-orthogonal postselection, i.e., $\langle
-|+\rangle \neq 0$, we can find that
\begin{equation}
\langle \hat{q}\rangle =\frac{2\varepsilon \eta }{\varepsilon ^{2}+\eta ^{2}}
\sigma  \label{kw}
\end{equation}
and
\begin{equation}
\langle \hat{p}\rangle =0.  \label{kv}
\end{equation}
When $\varepsilon =\pm \eta $ we will have the largest displacement $\pm
\sigma $ in position space and when $\varepsilon =0$, indicating that the
post-selected state of the system is absolutely orthogonal to the initial
state of the system, i.e., $\langle -|+\rangle =0$, the displacement of
pointer position is zero. However, the displacement of the pointer is always
zero in momentum space.

\end{document}